\documentclass[11pt]{article} % use larger type; default would be 10pt
\usepackage{amsmath}
\usepackage{amsxtra}
\usepackage{amsfonts}
\usepackage[utf8]{inputenc} % set input encoding (not needed with XeLaTeX)

\usepackage{geometry} % to change the page dimensions
\usepackage{graphicx} % support the \includegraphics command and options

\usepackage{booktabs} % for much better looking tables
\usepackage{array} % for better arrays (eg matrices) in maths
\usepackage{paralist} % very flexible & customisable lists (eg. enumerate/itemize, etc.)
\usepackage{verbatim} % adds environment for commenting out blocks of text & for better verbatim
\usepackage{subfig} % make it possible to include more than one captioned figure/table in a single float
\usepackage{fancyhdr} % This should be set AFTER setting up the page geometry
\usepackage{sectsty}
\allsectionsfont{\sffamily\mdseries\upshape} % (See the fntguide.pdf for font help)
\usepackage[nottoc,notlof,notlot]{tocbibind} % Put the bibliography in the ToC
\usepackage[titles,subfigure]{tocloft} % Alter the style of the Table of Contents

\title{Thermal conductivity for a general bidimensional dilute gas within
the Chapman-Enskog approximation}

\author{A. R. M\'endez\footnote{amendez@correo.cua.uam.mx}, A. L. Garc\'{i}a-Perciante\footnote{algarcia@correo.cua.uam.mx} and E. S. Escobar-Aguilar\footnote{e.ezkovar@gmail.com}}
\date{\today} 

\begin{document}
\maketitle
\abstract{In this work we explicitly calculate the thermal conductivity for a general bidimensional dilute gas of neutral molecules by solving Boltzmann's 
equation. Chapman-Enskog's method is used in order to analytically obtain this transport coefficient to first approximation in terms of a 
collision integral for an unspecified molecular interaction model. In the particular case of hard disks interactions, the result is shown to be 
consistent with previous work by J. V. Sengers \cite{Sengers}, yielding the expected $T^{1/2}$ dependence with the temperature. This dependence is 
widely used for dense gases as the low density limit in the Enskog expansion. The case of Maxwellian molecules is also explored where a linear 
dependence with the temperature is obtained.}
\section*{}
Keywords: Thermal conductivity; diluted gases; bidimensional systems; Chapman-Enskog method.\\
PACS: 51.10.+y; 51.20.+d.

\section{Introduction}

Most theoretical work on the field of bidimensional gaseous systems
is limited to the case of dense gases. Moreover, the question of whether
a density expansion is appropriate and the origin of the divergence
of the corresponding transport coefficients for low density limits
have been topics of intense research \cite{ Sengers,Sengers1,Gass,Risso1,Risso2,Cohen,Garcia-Rojo, Hanley,Ernst}.
To the authors' knowledge, the only work in which transport coefficients
for the dilute bidimensional gas are calculated can be found in Ref.
\cite{Sengers}. In that work, only the particular case of a hard
disks gas is presented as a side calculation in order to address the
triple collision contribution in such a system.  

In the present work an explicit calculation of the thermal conductivity
for a dilute bidimensional gas is developed in detail without specifying
a particular interaction potential. The heat flux and corresponding
conductivity are then obtained for two particular molecular models,
hard disks and Maxwellian particles, the former yielding results consistent
with Ref. \cite{Sengers}. It is worthwile to point out that these resultas cannot 
be directly deduced as a particular case of the 3D one. In particular, the dependence
of the scattering cross-section with the scattering angle cannot be obtained as a reduction of the 
corresponding in three dimensions. These expressions can be introduced in
bidimensional numerical simulations and are feasible of generalizations,
for example to the relativistic scenario. Additionally, the formalism
here presented in detail, can be useful in the calculation of other
transport coefficients.

The rest of this work is organized as follows. In Section 2 we briefly
review the basic formalism for calculating the heat flux from the
linearized Boltzmann equation within the Chapman-Enskog approximation,
here in the case of a two dimensional gas. The general expressions
for the heat flux and thermal conductivity are obtained in Sect. 3
in terms of a collision integral which is calculated to some detail
in the Appendix. Section 4 addresses the particular cases of hard
disks and Maxwellian molecules. The final section includes a brief
discussion and some final remarks.

\section{Boltzmann equation and Chapman-Enskog approximation}

The evolution equation for the single particle distribution function,
corresponding to a dilute bidimensional gas of neutral non-interacting
molecules is given by the Boltzmann equation \cite{Ch-E} 
\begin{equation}
\frac{\partial f}{\partial t}+\vec{v}\cdot \nabla f=\int\left[f'f'_{1}-ff_{1}\right]g\sigma d\chi d^{2}v_{1}.\label{1}
\end{equation}
Here $f$, the distribution function, corresponds to the particle
density in the four dimensional phase space such that $f\left(\vec{r},\vec{v},t\right)d^{2}rd^{2}v$
yields the number of particles at time $t$ with a position vector
between $\vec{r}$ and $\vec{r}+d\vec{r}$ and velocity between $\vec{v}$
and $\vec{v}+d\vec{v}$. The right hand side of Eq. (\ref{1}) accounts
for changes in the single particle distribution function due to collisions.
Non-primed and primed quantities denote values before and after the
collision respectively. The molecule's relative velocity is denoted
by $\vec{g}$, with $g$ being its magnitude. Also $\chi$ and $\sigma$
are the 2D counterparts to the solid angle and
the scattering cross section in the 3D case respectively. However it is worthwhile to mention at this point that $\sigma$ (which has dimensions of length) depends on $g$ and $\chi$ even for a hard disks model.

The local equilibrium solution of the Boltzmann equation for a bidimensional
system can be shown to be given by 
\begin{equation}
f^{\left(0\right)}=\frac{nm}{2\pi kT}e^{-\frac{mc^{2}}{2kT}},\label{2}
\end{equation}
by means of which one can establish the corresponding expressions
for the state variables, in this case the particle density 
\begin{equation}
n=\int fd^{2}v,\label{3}
\end{equation}
the hydrodynamic velocity 
\begin{equation}
n\vec{u}=\int f\vec{v}d^{2}v,\label{4}
\end{equation}
and the internal energy density
\begin{equation}
\varepsilon=\int f\left(\frac{1}{2}mv^{2}\right)d^{2}v.\label{5}
\end{equation}
The relationship between the molecular ($\vec{v}$), hydrodynamic ($\vec{u}$)
and peculiar ($\vec{c}$) velocities appearing in the previous
expressions is given by $\vec{v}=\vec{u}+\vec{c}$ being $\vec{v}$
the velocity of a molecule as measured by an arbitrary observer while
$\vec{c}$ is the one seen from a system comoving with the fluid,
that is an observer moving with velocity $\vec{u}$.

At this point it is convenient to introduce the Chapman-Enskog expansion.
Such a procedure allows one to express the complete solution to Eq.
(\ref{1}) as an infinite series of terms ordered in the Knudsen parameter.
This parameter quantifies the deviation from the local equilibrium distribution
function due to the gradients present in the system \cite{Ch-E}.
To first order in the gradients, corresponding to the Navier-Stokes
regime, one thus has 
\begin{equation}
f=f^{(0)}(1+\phi),\label{6}
\end{equation}
In this context, as is well known, the balance equations for the variables given in Eqs. (\ref{3})-(\ref{5}),
can be readily established by multiplying Eq. (\ref{1}) by a collisional
invariant and integrating over velocity space. Such a set requires
closure, or constitutive, relations expressing the dissipative fluxes
in terms of gradients of the state variables. In particular, for the
heat flux one obtains 
\begin{equation}
\vec{q}=\int f\phi\left(\frac{1}{2}mc^{2}\right)\vec{c}d^{2}c,\label{7}
\end{equation}
which will be used in this work to obtain an analytic expression for
the thermal conductivity $\kappa$ from the Fourier relation $\vec{q}=-\lambda\nabla T$.

The solution for $\phi$ is obtained from the linearized Boltzmann
equation to first order in Knudsen's parameter which is given by \cite{Ch-E}
\begin{equation}
\frac{df^{\left(0\right)}}{dt}=\mathcal{I}\left(\phi\right),\label{8}
\end{equation}
where the linearized collision kernel reads 
\begin{equation}
\mathcal{I}\left(\phi\right)=\int f^{\left(0\right)}f_{1}^{\left(0\right)}\left[\phi_{1}'+\phi'-\phi_{1}-\phi\right]g\sigma d\chi dv_{1}^{2}.\label{9}
\end{equation}
Following the usual steps in the Chapman-Enskog method \cite{Ch-E},
the vector part of Eq. (\ref{8}) can be shown to be given in this
case by 
\begin{equation}
f^{\left(0\right)}\left(\frac{mc^{2}}{2kT}-2\right)\vec{c}\cdot\frac{\nabla T}{T}=\mathcal{I}\left(\phi\right),\label{11}
\end{equation}
where we have neglected the terms involving the gradient of the hydrodynamic
velocity since it will not be coupled with the heat flux.

\section{Solution to the integral equation}

The Chapman-Enskog hypothesis together with the linearized Boltzmann
equation yield the integral equation (\ref{11}). Although
it is much simpler to solve than the integro-differential equation
 (\ref{1}), its complete solution can be involved. However,
as will be shown, the two-dimensional calculation is straightforward
and the collisional bracket corresponding to the heat flux can be
calculated directly for a hard disks collisional model.

The solution to Eq. (\ref{11}), following the same procedure as in
the 3D case, can be written as 
\begin{equation}
\phi=\mathcal{A}\left(c\right)\vec{c}\cdot\frac{\nabla T}{T},\label{17}
\end{equation}
once the subsidiary conditions are enforced \cite{Ch-E}. The function $\mathcal{A}\left(c\right)$
needs to be determined such that it satisfies the inhomogeneous equation
(\ref{11}), being still subject to the following condition, which
also arises from the subsidiary conditions, 
\begin{equation}
\frac{\nabla T}{T}\cdot\int f^{\left(0\right)}\mathcal{A}\left(c\right)\vec{c}\vec{c}d^{2}c=\frac{1}{2}\frac{\nabla T}{T}\cdot\mathbb{I}
\int f^{\left(0\right)}\mathcal{A}\left(c\right)c^{2}d^{2}c=\vec{0},\label{18}
\end{equation}
where $\mathbb{I}$ is the identity matrix. Introducing Eq. (\ref{17})
in Eq. (\ref{11}) and defining an auxiliary variable $x=mc^{2}/2kT$
, the integral equation for $\mathcal{A}\left(x\right)$ can be written
as 
\begin{equation}
f^{\left(0\right)}\left(x-2\right)\vec{c}\cdot\frac{\nabla T}{T}=\mathcal{I}\left(\mathcal{A}\left(x\right)\vec{c}\cdot\frac{\nabla T}{T}\right)\label{19}
\end{equation}
where $\mathcal{A}\left(x\right)$ satisfies 
\begin{equation}
\int\mathcal{A}\left(x\right)p\left(x\right)dx=0,\label{20}
\end{equation}
with 
\begin{equation}
p\left(x\right)=xe^{-x},\label{21}
\end{equation}
being the weight function. Following the standard procedure, $\mathcal{A}\left(x\right)$
is written as an infinite series in orthogonal polynomials. In this
case, as in the three dimensional one, the appropriate basis are the
Sonine polynomials $\mathcal{L}_{n}^{\left(k\right)}$. Taking into
account Eqs. (\ref{19})-(\ref{21}), the appropriate polynomials
for the 2D case correspond to $k=1$ since 
\begin{equation}
\int e^{-x}\mathcal{L}_{n}^{\left(1\right)}\left(x\right)\mathcal{L}_{m}^{\left(1\right)}\left(x\right)xdx=\frac{\Gamma\left(n+2\right)}{n!}\delta_{mn},\label{22}
\end{equation}
with 
\begin{equation}
\mathcal{L}_{0}^{\left(1\right)}\left(x\right)=1\quad\text{and}\quad\mathcal{L}_{1}^{\left(1\right)}\left(x\right)=2-x.\label{23}
\end{equation}
Thus, we have 
\begin{equation}
\mathcal{A}\left(x\right)=\sum_{n=0}^{\infty}a_{n}\mathcal{L}_{n}^{\left(1\right)}\left(x\right),\label{24}
\end{equation}
and the condition in Eq. (\ref{20}) is written as 
\begin{equation}
\sum_{n=0}^{\infty}a_{n}\int p\left(x\right)\mathcal{L}_{n}^{\left(1\right)}\left(x\right)dx=0,\label{25}
\end{equation}
which implies $a_{0}=0$. Thus, the solution to the integral equation
(\ref{11}) is given by Eq. (\ref{17}) with $\mathcal{A}\left(x\right)=\sum_{n=1}^{\infty}a_{n}\mathcal{L}_{n}^{\left(1\right)}\left(x\right)$
and $a_{n}$ obtained from 
\begin{equation}
-f^{\left(0\right)}\mathcal{L}_{1}^{\left(1\right)}\left(x\right)\vec{c}=\sum_{n=1}^{\infty}a_{n}\mathcal{I}\left(\mathcal{L}_{n}^{\left(1\right)}\left(x\right)\vec{c}\right).\label{27}
\end{equation}
With the proposed solution, the heat flux can be written as

\begin{equation}
\vec{q}=\frac{n}{m}\left(kT\right)^{2}\frac{\nabla T}{T}\sum_{n=1}^{\infty}a_{n}\int\mathcal{L}_{n}^{\left(1\right)}\left(x\right)xp\left(x\right)dx.\label{30}
\end{equation}
Writing $x$ in terms of the Sonine polynomials as $x=2\mathcal{L}_{0}^{\left(1\right)}\left(x\right)-\mathcal{L}_{1}^{\left(1\right)}\left(x\right)$, and using the orthogonality condition (\ref{22}) one obtains that
\begin{equation}
\vec{q}=-2\frac{n}{m}\left(kT\right)^{2}\frac{\nabla T}{T}a_{1}.\label{32}
\end{equation}
Thus, only $a_{1}$ is required for the purposes of this work, the
scalar product of Eq. (\ref{27}) with $\mathcal{L}_{m}^{\left(1\right)}\left(x\right)\vec{c}$ is calculated and the resulting equation is integrated over velocity
space. The left hand side yields 
\begin{equation}
\frac{2nkT}{m}\int\mathcal{L}_{1}^{\left(1\right)}\left(x\right)\mathcal{L}_{m}^{\left(1\right)}\left(x\right)p\left(x\right)dx=4\frac{nkT}{m}\delta_{1m},\label{33}
\end{equation}
and thus 
\begin{equation}
4\frac{nkT}{m}\delta_{1m}=-\sum_{n=1}^{\infty}a_{n}\int\mathcal{L}_{m}^{\left(1\right)}\left(x\right)\vec{c}\cdot\mathcal{I}\left(\mathcal{L}_{n}^{\left(1\right)}\left(x\right)\vec{c}\right)d^{2}c.\label{34}
\end{equation}
Equation (\ref{34}) can now be solved for $a_{1}$ to different orders
of approximation using the standard variational method, whose details
can be found in Ref. \cite{Hirshfelder}. To first approximation one
obtains 
\begin{equation}
a_{1}=-4\frac{nkT}{m}\left(\int\mathcal{L}_{1}^{\left(1\right)}\left(x\right)\vec{c}\cdot\mathcal{I}\left(\mathcal{L}_{1}^{\left(1\right)}\left(x\right)\vec{c}\right)d^{2}c\right)^{-1}.\label{35}
\end{equation}
The integral above is usually written in terms of a collisional bracket
and requires, in most cases, tedious calculations in order to be evaluated.
However, using momentum conservation, one can show that
\begin{eqnarray}
\mathcal{I}\left(\mathcal{L}_{1}^{\left(1\right)}\left(x\right)\vec{c}\right) 
 & = & -\frac{m}{kT}\mathcal{I}\left[\frac{c^{2}}{2}\vec{c}\right],\label{36}
\end{eqnarray}
and by interchanging particles after and before the collision
as well as dummy indices, 
\begin{equation}
\int\vec{H}\cdot\mathcal{I}\left(\vec{G}\right)d^{2}c=\int\vec{G}\cdot\mathcal{I}\left(\vec{H}\right)d^{2}c.\label{37}
\end{equation}
Using Eqs. (\ref{36}) and (\ref{37}) one obtains 
\begin{equation}
\int\mathcal{L}_{1}^{\left(1\right)}\left(x\right)\vec{c}\cdot\mathcal{I}\left(\mathcal{L}_{1}^{\left(1\right)}\left(x\right)\vec{c}\right)d^{2}c=\frac{m}{kT}I,\label{38-1}
\end{equation}
where 
\begin{equation}
I=\int\mathcal{I}\left[\left(\frac{mc^{2}}{2kT}-2\right)\vec{c}\cdot\right]\frac{c^{2}}{2}\vec{c}d^{2}c,\label{39}
\end{equation}
and thus $a_{1}$ can be expressed as 
\begin{equation}
a_{1}=-4\frac{nk^{2}T^{2}}{m^{2}}I^{-1}.\label{40}
\end{equation}
The detailed calculation of $I$ is shown in the Appendix where the
expression 
\begin{equation}
I=-\frac{n^{2}\beta}{32\pi}\int\exp\left(-\frac{\beta}{2}g^{2}\right)g^{5}\sin^{2}\chi\sigma\left(\chi,g\right)d\chi d^{2}g,\label{eq:41}
\end{equation}
is obtained with $\beta=m/2kT$. It is worthwhile to point out that
in the bidimensional case, this integral is in principle simpler
to calculate than in the 3D case. The general expression for
the thermal conductivity of a dilute bidimensional gas is thus given
by 
\begin{equation}
\lambda=-\frac{1}{I}\frac{8n^{2}k^{4}T^{3}}{m^{3}}.\label{eq:45}
\end{equation}
This expression for the transport coefficient is the main result of
this work. In the following section, two particular models for the
intermolecular interaction will be introduced in order to obtain explicit
temperature dependence of $\lambda$.

\section{The hard disks and Maxwellian molecules models}

In order to perform the integration in Eq. (\ref{eq:41}) with respect
to the scattering angle $\chi$, a particular model for the 2D scattering cross section 
needs to be introduced. For example, in the case of
hard disks $\sigma\left(\chi,g\right)$ depends on the scattering
angle as follows 
\begin{equation}
\sigma=\left|\frac{db}{d\chi}\right|=\left|\sin\left(\frac{\chi}{2}\right)\right|,
\end{equation}
where $b$ is the impact parameter, this leads to the following expressions for $I$ and $a_{1}$ (see
Appendix): 
\begin{equation}
I=-n^{2}d\beta^{-5/2}\sqrt{\frac{\pi}{2}}\label{eq:42}
\end{equation}
and

\begin{equation}
a_{1}=\frac{1}{nd}\left(\frac{m}{\pi kT}\right)^{1/2},
\end{equation}
respectively.
Using the results above, the heat flux can be written as 
\begin{equation}
\vec{q}=-\frac{2}{d}\left(\frac{k^{3}T}{\pi m}\right)^{1/2}\nabla T,
\end{equation}
from which one obtains the thermal conductivity for the hard disk dilute
gas as
\begin{equation}
\lambda=\frac{2}{d}\left(\frac{k^{3}T}{\pi m}\right)^{1/2}.\label{43}
\end{equation}
consistently with the results quoted in Ref. \cite{Sengers}.

The Maxwellian disks model is a particular case of the point centers
of repulsion potential and approximates the molecular interaction
for the case of high temperature gases in which the repulsive forces
dominate over the attractive ones. In such a case, the potential can
be expressed as \cite{Hirshfelder} 
\begin{equation}
\phi\left(r\right)=\frac{\kappa}{\nu-1} r^{-(\nu-1)},
\end{equation}
where $\nu$ is a measure of the hardness of the molecules. Maxwellian
molecules are characterized by $g\sigma\left(\chi,g\right)$ being
independent of $g$, a property that in two dimensions is obtained
for $\nu=3$. In such a case one obtains a scattering angle given
by 
\begin{equation}
\chi=\pi\left(1-\left(1+\frac{2\kappa}{m}\frac{1}{g^{2}b^{2}}\right)^{-1/2}\right),\label{43.1}
\end{equation}
which leads to the following expression for $\sigma(g,\chi)$
\begin{equation}
\sigma\left(g,\chi\right)=\sqrt{\frac{2\kappa}{m}}\frac{\pi^{2}}{g}\left(\chi\left(2\pi-\chi\right)\right)^{-3/2}.\label{43.2}
\end{equation}
With this result, the integral $I$ yields 
\begin{equation}
I=-\sqrt{\frac{2\kappa}{m}}\frac{\pi^{2}n^{2}}{2\beta^{2}}\Gamma,
\end{equation}
where $\Gamma$ is given in by Eq. (\ref{eq:66-b}) in the Appendix.
Thus the heat flux and thermal conductivity are given by 
\begin{equation}
\vec{q}=-\frac{1}{\Gamma}\frac{4}{\pi^2}\sqrt{\frac{1}{2m\kappa}}k^{2}T\nabla T\label{eq:67-b}
\end{equation}
and 
\begin{equation}
\lambda=\frac{1}{\Gamma}\frac{4}{\pi^{2}}\sqrt{\frac{1}{2m\kappa}}k^{2}T
\end{equation}
respectively, with $\Gamma\approx0.212383$. It should de emphasized
that for this model, the thermal conductivity grows linearly with
the temperature, which is analogous to result in the 3D case.

\section{Discussion and final remarks}

In this work, the standard procedure to establish transport coefficients
has been used in order to obtain the thermal conductivity for a dilute
bidimensional gas. As mentioned above, the temperature dependence
$T^{1/2}$ and the proportionality constant for the hard disks case
are consistent with the result in Ref. \cite{Sengers}, which is widely
used as the low density limit when considering dense gases. 

The main contribution of this work, the thermal conductivity given
by Eqs. (\ref{eq:41}) and (\ref{eq:45}), allows the introduction
of any molecular model in order to obtain numerical approximations
for this coefficient in bidimensional dilute gases. Additionally,
the thermal conductivity is here calculated for Maxwellian molecules
where the temperature dependence coincides with the three dimensional
case. It is of the authors' opinion that the explicit and systematic
fashion in which this work is presented will be useful for generalizations
to other dimensions, intermolecular potentials and transport coefficients. To the
authors' knowledge, this type of calculation has not been published
in the available literature and could also be used as a baseline for
the calculation of bidimensional transport coefficients for relativistic
dilute gases which will be reported elsewhere.

%For acknowledgements section, please don't number the section, please begin it with \section*{Acknowledgements}
\section*{Acknowledgments} The authors acknowledge financial support from CONACyT through grant
number CB2011/167563. 

\appendix
%dummy comment inserted by tex2lyx to ensure that this paragraph is not empty
\section*{Appendix} \label{Appendix} 

In this Appendix, the integral $I$ given in Eq.
(\ref{39}), is calculated following the procedure described in Ref.
\cite{Kremer}. Firstly, a change of variables is introduced (to relative
and center of mass velocities): 
\begin{equation}
\vec{g}=\vec{c}-\vec{c}_{1}\qquad\vec{G}=\vec{c}_{1}+\vec{c},\label{44}
\end{equation}
such that 
\begin{equation}
\vec{c}=\frac{\vec{G}+\vec{g}}{2}\qquad\vec{c}_{1}=\frac{\vec{G}-\vec{g}}{2},\label{45}
\end{equation}
and the corresponding Jacobian is $J=2^{-2}$. By introducing these new
variables in Eq. (\ref{39}), $I$ can be expressed as 
{\small
\begin{equation}
I=\frac{1}{32}\int f^{\left(0\right)}f_{1}^{\left(0\right)}\left(\frac{m}{2kT}\frac{\left(G^{2}+2\vec{G}\cdot\vec{g}+g^{2}\right)}{4}-2\right)\left(\vec{G}+\vec{g}\right)\cdot\left[\vec{G}\cdot\left(\vec{g}'\vec{g}'-\vec{g}\vec{g}\right)\right]g\sigma d\Omega d^{2}Gd^{2}g.\label{46}
\end{equation}}
Introducing the parameter $\beta=m/2kT$ , and the equilibrium distribution
function 
\begin{equation}
f^{\left(0\right)}=\left(\frac{n}{\pi}\beta\right)e^{-\beta c^{2}},\label{47}
\end{equation}
one has 
\begin{eqnarray}
I=& \frac{1}{32}\left(\frac{n}{\pi}\beta\right)^{2}\int\exp\left[-\frac{\beta}{2}\left(G^{2}+g^{2}\right)
\right]\left[\beta\left(\frac{G^{2}+2\vec{G}\cdot\vec{g}+g^{2}}{4}\right)-2\right]\left(\vec{G}+\vec{g}\right)\nonumber\\
 & \left[\vec{G}\cdot\left(\vec{g}'\vec{g}'-\vec{g}\vec{g}\right)\right]g\sigma\left(\chi,g\right)d\chi d^{2}Gd^{2}g.\label{48}
\end{eqnarray}
The integration with respect to $\vec{G}$ can be separated into two
terms as follows 
{\small
\begin{eqnarray}
\int\exp\left[-\frac{\beta}{2}G^{2}\right]\left[\beta\left(\frac{G^{2}+2\vec{G}\cdot\vec{g}+g^{2}}{4}\right)-2\right]\left(\vec{G}+\vec{g}\right)\cdot\left[\vec{G}\cdot\left(\vec{g}'\vec{g}'
-\vec{g}\vec{g}\right)\right]g\sigma\left(\chi,g\right)d\chi d^{2}Gd^{2}g \nonumber \\
=\int\exp\left[-\frac{\beta}{2}G^{2}\right]\left[\beta\left(\frac{G^{2}+2\vec{G}\cdot\vec{g}+g^{2}}{4}\right)-2\right]\vec{G}\cdot\left[\vec{G}\cdot\left(\vec{g}'\vec{g}'-\vec{g}\vec{g}\right)\right]g\sigma\left(\chi,g\right)d\chi d^{2}Gd^{2}g\label{49}\\
+\int\exp\left[-\frac{\beta}{2}G^{2}\right]\left[\beta\left(\frac{G^{2}+2\vec{G}\cdot\vec{g}+g^{2}}{4}\right)-2\right]\vec{g}\cdot\left[\vec{G}\cdot\left(\vec{g}'\vec{g}'-\vec{g}\vec{g}\right)\right]g\sigma\left(\chi,g\right)d\chi d^{2}Gd^{2}g.\nonumber 
\end{eqnarray}}
For the first term (second line in Eq. (\ref{49})), since 
\[\vec{G}\cdot\left[\vec{G}\cdot\left(\vec{g}'\vec{g}'-\vec{g}\vec{g}\right)\right]=\sum_{i,j=1}^{2}G_{i}G_{j}\left(g_{i}'g_{j}'-g_{i}g_{j}\right),\]
one has
{\small 
\begin{eqnarray}
\sum_{i,j=1}^{2}\left(g_{i}'g_{j}'-g_{i}g_{j}\right)\int\exp\left[-\frac{\beta}{2}G^{2}\right]\left[\beta\left(\frac{G^{2}+2\vec{G}\cdot\vec{g}+g^{2}}{4}\right)-2\right]G_{i}G_{j}d^{2}G \nonumber\\
=\sum_{i,j=1}^{2}\left(g_{i}'g_{j}'-g_{i}g_{j}\right)\int\exp\left[-\frac{\beta}{2}G^{2}\right]\left[\beta\frac{\left(G^{2}+g^{2}\right)}{4}-2\right]G_{i}G_{j}d^{2}G,\label{50} 
\end{eqnarray}}
which vanishes since the integral is proportional to $\delta_{ij}$
and 
\begin{equation}
\left(g_{i}'g_{j}'-g_{i}g_{j}\right)\delta_{ij}=g'^{2}-g^{2}=0.\label{51}
\end{equation}
For the second term (third line in Eq. (\ref{49})), considering that
\[\vec{g}\cdot\left[\vec{G}\cdot\left(\vec{g}'\vec{g}'-\vec{g}\vec{g}\right)\right]=\sum_{i,j=1}^{2}g_{i}G_{j}\left(g_{i}'g_{j}'-g_{i}g_{j}\right),\]
the integration with respect to $\vec{G}$ yields 
{\small
\begin{eqnarray}
& &\sum_{i,j=1}^{2}g_{i}\left(g_{i}'g_{j}'-g_{i}g_{j}\right)g\int\exp\left[-\frac{\beta}{2}G^{2}\right]\left[\beta\left(\frac{G^{2}+2\vec{G}\cdot\vec{g}+g^{2}}{4}\right)-2\right]G_{j}d^{2}G\nonumber \\
&=&\sum_{i,j=1}^{2}g_{i}\left(g_{i}'g_{j}'-g_{i}g_{j}\right)g\int\exp\left[-\frac{\beta}{2}G^{2}\right]\left[\beta\left(\frac{G^{2}+2\vec{G}\cdot\vec{g}+g^{2}}{4}\right)-2\right]G_{j}d^{2}\label{eq:4448}\\
&=&\frac{\pi g}{\beta}\sum_{i,j=1}^{2}g_{i}\left(g_{i}'g_{j}'-g_{i}g_{j}\right)g_{j} \hspace{7.05cm} \nonumber 
\end{eqnarray}}
and thus 
\begin{equation}
I=\frac{n^{2}\beta}{32\pi}\sum_{i,j=1}^{2}\int\exp\left(-\frac{\beta}{2}g^{2}\right)\left(g_{i}'g_{j}'-g_{i}g_{j}\right)g_{i}g_{j}g\sigma\left(\chi,g\right)d\chi d^{2}g,\label{53}
\end{equation}
The integration over the scattering angle $\chi$ can be performed
by aligning the $y$ axis with the initial relative velocity such
that 
\begin{equation}
\vec{g}=g\left(0,1\right),\label{54}
\end{equation}
and 
\begin{equation}
\vec{g}'=g\left(\sin\chi,\cos\chi\right).\label{55}
\end{equation}
Since 
\begin{equation}
\int_{0}^{2\pi}\sigma\left(g,\chi\right)\left[g_{i}'g_{j}'-g_{i}g_{j}\right]g_{i}g_{j}d\chi=-g^{4}\int_{0}^{2\pi}\sigma\left(g,\chi\right)\sin^{2}\chi d\chi\label{eq:56-b}
\end{equation}
we have 
\begin{equation}
I=-\frac{n^{2}\beta}{32\pi}\int\exp\left(-\frac{\beta}{2}g^{2}\right)g^{5}\sin^{2}\chi\sigma\left(\chi,g\right)d\chi d^{2}g,\label{53-2}
\end{equation}
which is Eq. (\ref{eq:41}).

A specific molecular interaction model needs to be introduced in order
to perform the $\chi$ integration. For hard disks $\sigma=\frac{d}{2}\left|\sin\left(\frac{\chi}{2}\right)\right|$,
and thus 
\begin{equation}
I=-\frac{d}{2}\frac{n^{2}\beta}{32\pi}\int\exp\left(-\frac{\beta}{2}g^{2}\right)g^{5}\left|\sin\left(\frac{\chi}{2}\right)\right|\sin^{2}\chi d\chi d^{2}g\label{eq:57-b}
\end{equation}
Integrating over the scattering angle $\int_{0}^{2\pi}\left|\sin\left(\chi/2\right)\right|\sin^{2}\chi d\chi=32/15$,
one obtains 
\begin{equation}
I=-n^{2}d\beta^{-5/2}\sqrt{\frac{\pi}{2}},\label{eq:58-b}
\end{equation}
which is the expression used in the main text to obtain the thermal
conductivity for this model.

On the other hand, for Maxwellian molecules, since $\chi$ is given
by 
\begin{equation}
\chi=\pi-\int_{0}^{s_{max}}\frac{2ds}{\sqrt{1-s^{2}-\frac{2}{mg^{2}}\Phi\left(\frac{b}{s}\right)}}\label{eq:59-b}
\end{equation}
where $s=b/r$ and $s_{max}$ is the positive root of $1-s^{2}-2\Phi\left(b/s\right)/mg^2=0$,
one has 
\begin{equation}
\chi=\pi-\int_{0}^{\frac{1}{\alpha}}\frac{2ds}{\sqrt{1-\alpha^{2}s^{2}}}=\pi\left(1-\frac{1}{\alpha}\right)\label{eq:60-b}
\end{equation}
where $\alpha^{2}=1+\left(2\kappa/m\right)\left(1/g^{2}b^{2}\right)$.
Equation (\ref{eq:60-b}) is the expression for $\chi$ given in Eq.
(\ref{43.1}). Solving for the impact parameter one obtains 
\begin{equation}
b\left(\chi\right)=\sqrt{\frac{2\kappa}{mg^2}}\left[\left(\frac{\pi}{\pi-\chi}\right)^{2}-1\right]^{-1/2}\label{eq:67-b-1}
\end{equation}
which leads to Eq. (\ref{43.2}) after differentiation. Introducing
$\sigma\left(g,\chi\right)$ in Eq. (\ref{eq:41}) yields

\begin{equation}
I=-\frac{n^{2}\pi^{2}}{2\beta^{2}}\sqrt{\frac{2\kappa}{m}}\Gamma,\label{53-2-1}
\end{equation}
where we have defined 
\begin{equation}
\Gamma=\int\sin^{2}\chi\left(\chi\left(2\pi-\chi\right)\right)^{-3/2}d\chi.\label{eq:66-b}
\end{equation}
The integral in equation (\ref{eq:66-b}) can be solved numerically
to obtain $\Gamma\approx0.212383$.
% You may incorporate your references as follows in your main tex file.
% Using BibTex is not recommended but can be handled.

\end{document}